\documentclass[paper,nofootinbib] {revtex4}

\usepackage{graphicx}
\usepackage{axodraw}
\usepackage{epsfig}
\usepackage{fancyhdr}
\usepackage{empheq}
\usepackage{mathbbol}
\usepackage{wasysym}
\usepackage{pstricks}
\usepackage{color}
\usepackage{bbm}
\def\yb{\ensuremath{Y_B}}
\def\qq{\mathbbmss{q}}

\begin{document}

\title{Charmed Baryonium}

\author{G Cotugno$^{\dag}$, R Faccini$^{\dag,\P}$,  AD Polosa$^\P$, C Sabelli$^{\dag,\P}$ } 
\affiliation{
$^\P$INFN Roma, Piazzale A. Moro 2, Roma, I-00185, Italy\\
$^\dag$Department of Physics, Universit\`a di Roma, `La Sapienza', Piazzale A. Moro 2, Roma, I-00185, Italy}

\begin{abstract}
We re-analyze the published data on the $Y(4630)\to  \Lambda_c \bar \Lambda_c$ and the $Y(4660)\to \psi(2S)\pi\pi$ with a consistent Ansatz and we find that the two observations are likely to be due to the {\it same state} $\yb$  with $M_{Y_B}=4660.7\pm 8.7$~MeV and $\Gamma_{Y_B}=61\pm23$~MeV. Under this hypothesis and reanalizing also the $e^+e^-\to J/\psi\pi\pi\gamma_{_{\rm ISR}}$ spectrum
we extract  ${\cal B}(\yb \to  \Lambda_c \bar \Lambda_c)/{\cal B}(\yb\to \psi(2S)\pi\pi)=25\pm 7$,  ${\cal B}(\yb \to  J/\psi\pi\pi)/{\cal B}(\yb\to \psi(2S)\pi\pi)<0.46$ @~90\% C.L., ${\cal B}(Y(4350)\to J/\psi\pi\pi)/{\cal B}(Y(4350)\to \psi(2S)\pi\pi)<3.4\times 10^{-3}$~@~90\% C.L.
and ${\cal B}(\yb \to  \psi(2S)\sigma)/{\cal B}(\yb\to \psi(2S)f_0)=2.0\pm 0.3$.  These conclusions strongly support the hypothesis of $\yb$ being the first observation of a charmed baryonium constituted by four quarks. From the analysis of the mass spectrum  we show that $Y(4350)$ and $Y_B$ are respectively consistent with the ground state and first radial excitation of the $\ell=1$ state. 
\end{abstract}

\maketitle

\thispagestyle{fancy}

{\bf \emph{Introduction}}.
The search of  the so called baryonia, exotic states originally thought to appear in nucleon-antinucleon systems, has 
a rather long and sometimes controversial history. Early work in this field can be found in~\cite{veneziano}
and quite recently there have been new interesting experimental indications as discussed 
{\it e.g.} in~\cite{nucleon}. 
In this letter we focus on higher mass scales, namely we refer to the hidden charm sector.  
We identify as a charmed baryonium a narrow structure, which we call  $\yb (J^{PC}=1^{--})$, with a decay pattern  
dominated by the $\Lambda_c\bar \Lambda_c$ baryon-antibaryon mode  and compare a set of other observables with the expectations of a tetraquark model.
Many different interpretations have been proposed in previous works on the nature of the $1^{--}$ neutral states in the 4~GeV region. The conventional charmonium assignation has been discussed in \cite{Ding:2007rg}. Besides  the charmonium hybrid interpretation \cite{4260hyb} and the {\it tetraquark} exlpanation \cite{Maiani:2005pe}, the possibility that $Y(4260)$ could be a threshold effect \cite{Rosner:2006vc} or a charm mesons molecule \cite{4260mol} have been considered too. Recently  \cite{Close:2009ag} and \cite{Guo:2008zg} proposed the $Y(4660)$ to be a $D^{*}\bar{D}_{1}$ and a $\psi^{'}f_{0}$ bound state respectively.

\begin{figure}
\begin{minipage}[t]{5truecm} 
\centering
\includegraphics[width=5.5truecm]{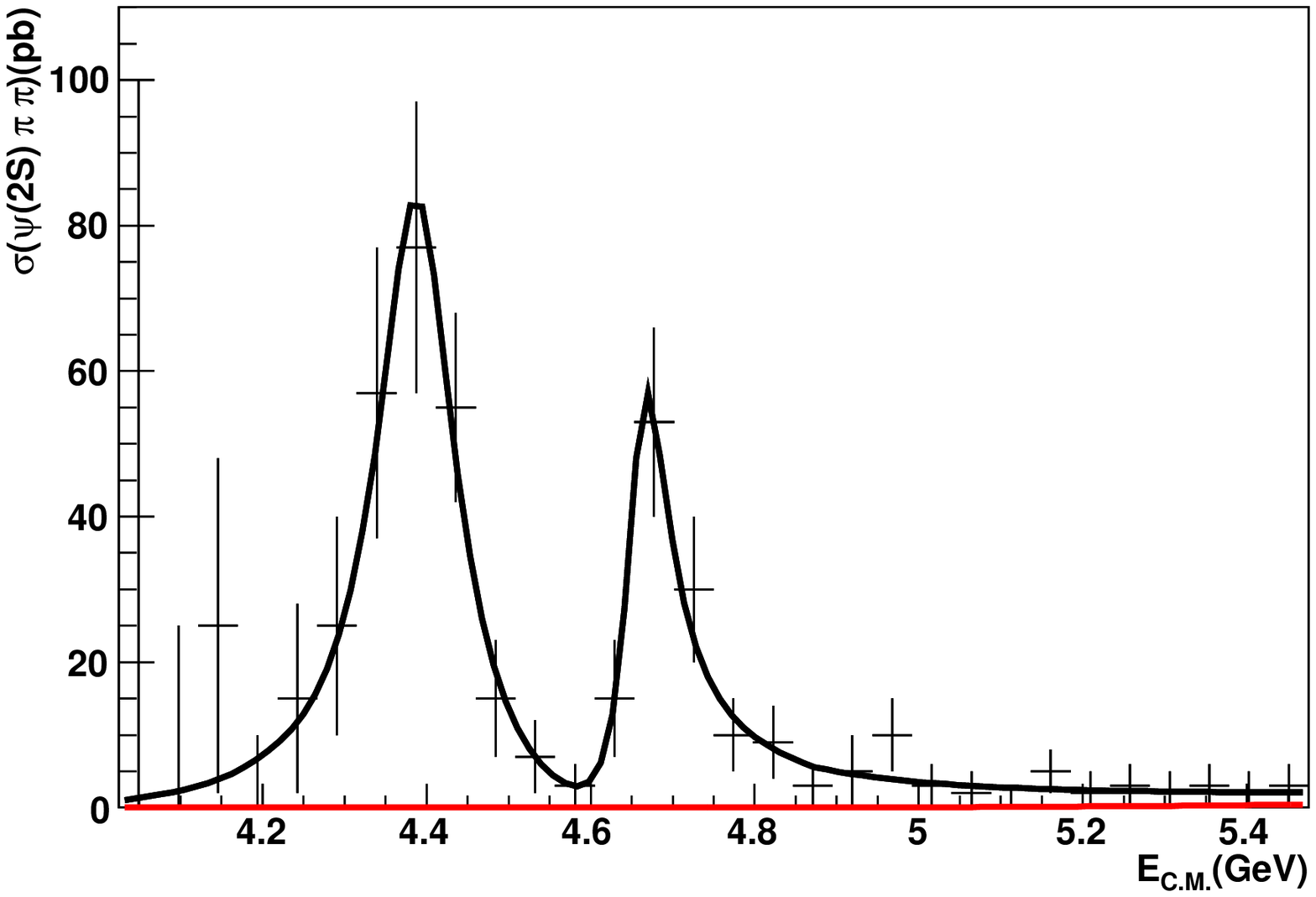}
\end{minipage}
\hspace{0.5truecm} 
\begin{minipage}[t]{5truecm}
\centering
\includegraphics[width=5.5truecm]{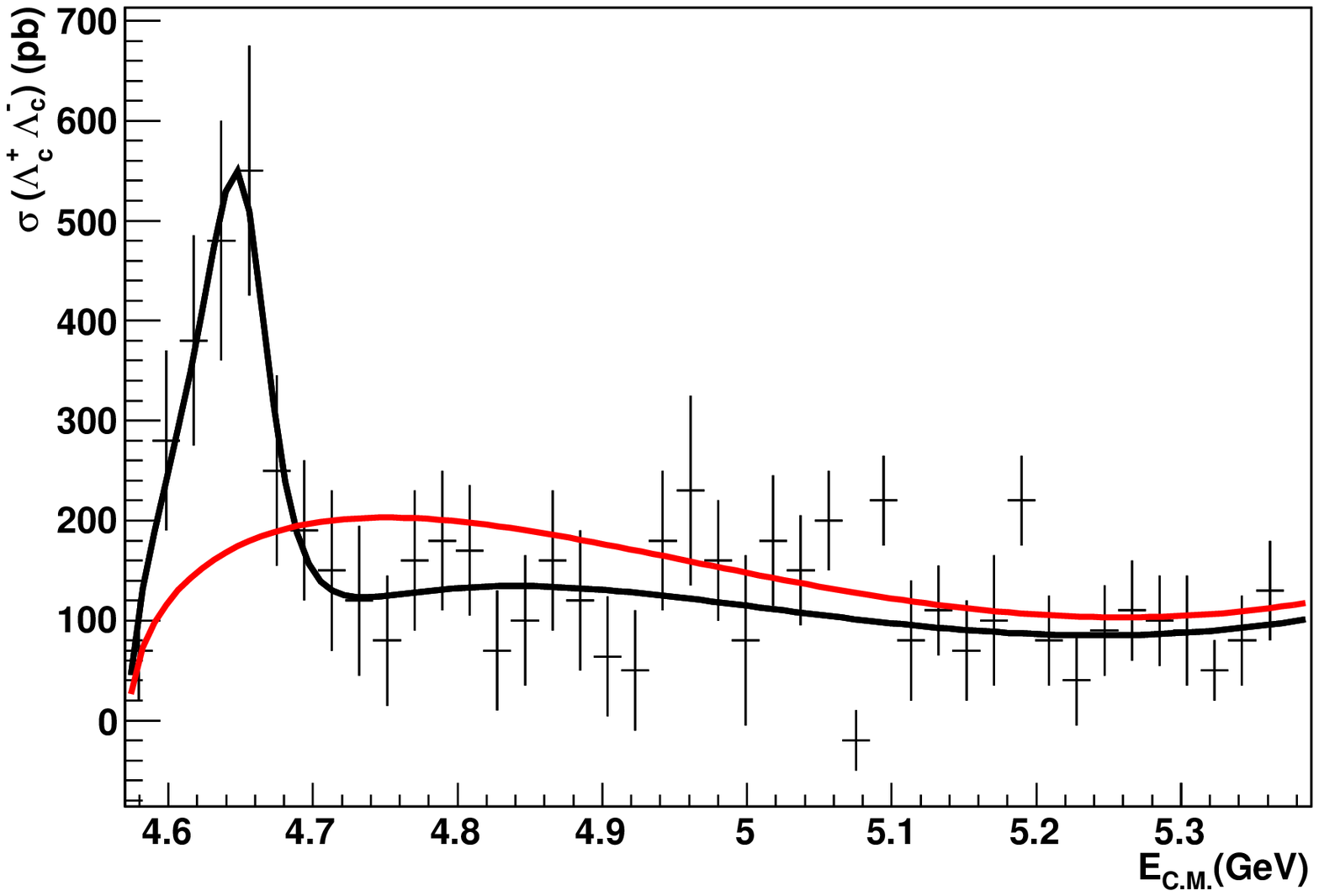}
\end{minipage}
\hspace{0.5truecm} 
\begin{minipage}[t]{5truecm}
\centering
\includegraphics[width=5.5truecm]{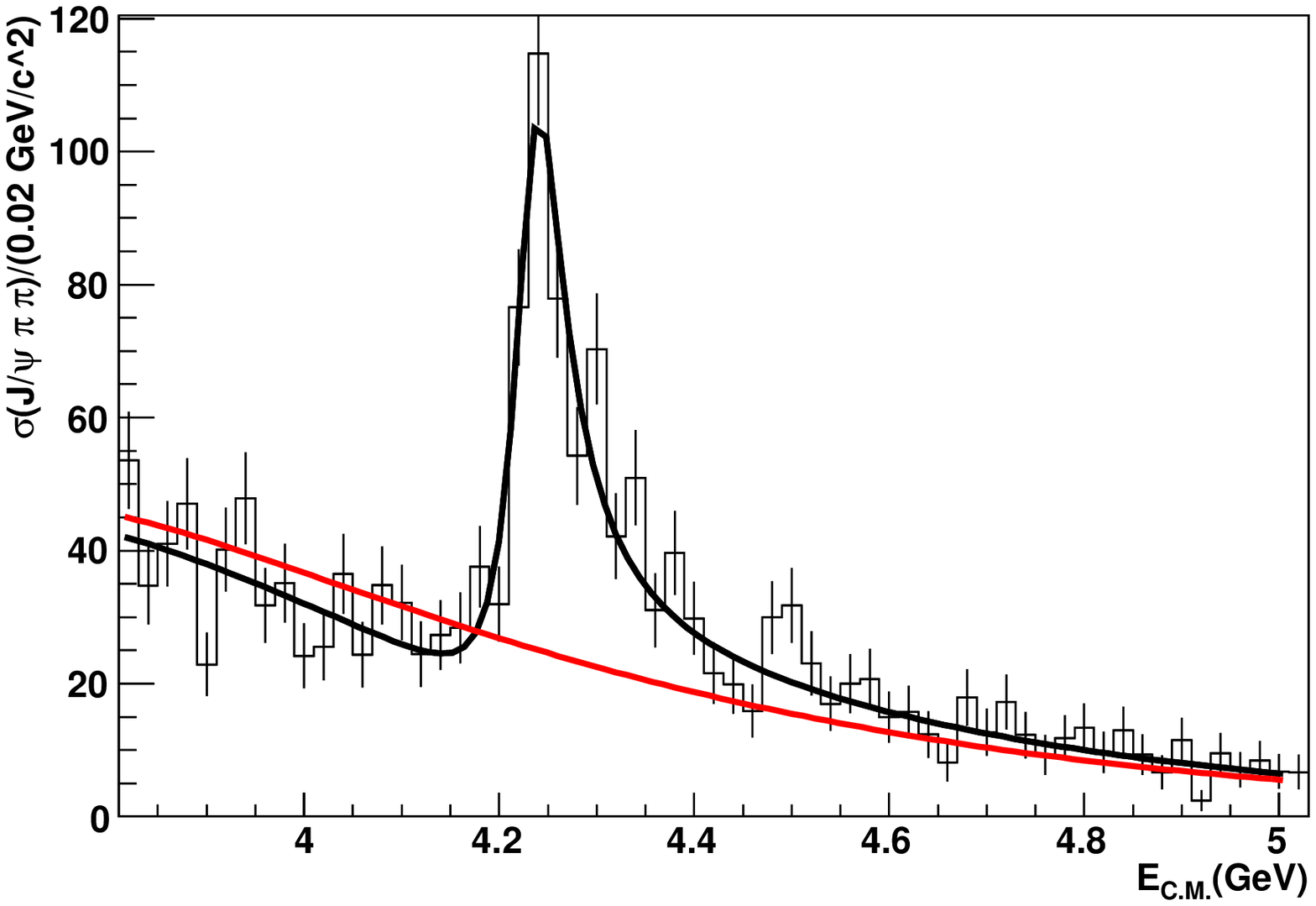}
\end{minipage}
\caption{
Fit of the cross section from the data of Belle Collaboration \cite{belle2,belle1} for the process $e^{+}e^{-}\rightarrow\psi(2S)\pi^{+}\pi^{-}$ (left) and $e^{+}e^{-}\rightarrow\Lambda^{+}_{c}\Lambda^{-}_{c}$ (center) and from the BaBar Collaboration~\cite{BaBar1}  for the process $e^{+}e^{-}\rightarrow J/\psi\pi^{+}\pi^{-}$ (right) . The black line represents the fit results, the red one shows the polynomial background.
}
\label{fig:tune}
\end{figure}

{\bf \emph{$Y_B$ from Belle data}}.
The existence of  $Y_B$  stems from our re-analysis of  data from the Belle experiment on the decay of the  $Y(4660)$ resonance 
 in $\psi(2S) \pi\pi$~\cite{belle2} and of  the 
 $Y(4630)$  in $\Lambda_c \bar \Lambda_c$~\cite{belle1}.  We fit the invariant mass spectra in Fig.~\ref{fig:tune}  with a binned likelihood, adopting a consistent signal model: a relativistic Breit-Wigner with comoving width, as detailed in Ref.~\cite{dre1}. Background is parameterized with a second order polynomial multiplied by the phase space.
 
The individual fits to the charmonium and baryonic  modes return  
$M_{Y_B}=4661\pm 9$~MeV,  $\Gamma_{Y_B}=61\pm23$~MeV with $\chi^{2}/{\rm d.o.f.}=7/20$ and
$M_{Y_B}=4661\pm 14$~MeV, $\Gamma_{Y_B}=63\pm23$~MeV with $\chi^{2}/{\rm d.o.f.}=51/35$, respectively.
The two results are consistent, strongly supporting the hypothesis that the two structures are evidences of the {\it same resonance}. In the following we will use the results of the first and more accurate, fit as reference.
From the same fits we also extract 
\begin{equation}
\frac{{\cal B}(Y_B\rightarrow\Lambda_{c}\bar \Lambda_{c})}{{\cal B}(Y_B\rightarrow\psi(2S)\pi^{+}\pi^{-})} = 25\pm 7,
\label{br}
\end{equation}
a result which highlights a strong affinity of the $Y_B$ to the baryon-antibaryon decay mode. It is to be noted that omitting interference, the ratio of the peak cross sections is about~11.

Another puzzling aspect of this meson is the absence of the decay mode into $J/\psi\pi^{+}\pi^{-}$. In order to try to interpret this observation, we also extract, with the same signal model, an upper limit on the data from BaBar on this decay (see Fig.~\ref{fig:tune}~\cite{BaBar1}) and extract,
with a bayesian approach,  ${\cal B}(\yb \to  J/\psi\pi\pi)/{\cal B}(\yb\to \psi(2S)\pi\pi)<0.46$ @~90\% C.L. The presence of the nearby $Y(4260)$ was properly accounted for.

By fitting consistently the $\psi(2S)\pi^+\pi^-$ spectrum in Fig.~\ref{fig:tune} we also extract the parameters of the $Y(4350)$ resonance. We obtain $M=4353\pm 13$~MeV and $\Gamma=118\pm 25$~MeV. We also find that  ${\cal B}(Y(4350) \to  J/\psi\pi\pi)/{\cal B}(Y(4350) \to \psi(2S)\pi\pi)< 3.4 \times 10^{-3}$ @~90\% C.L.

\begin{figure}[htb]
\centering
\vspace{-0.3truecm}
\includegraphics[width=5cm]{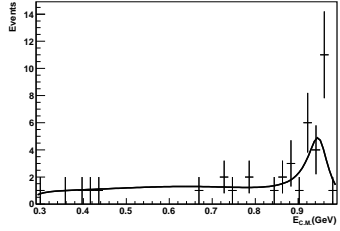}
\caption{
Fit of the dipionic mass distribution in the mass region for $\psi(2S)\pi^{+}\pi^{-}$: $4.0~ {\rm GeV} < m_{\pi^{+}\pi^{-}\psi(2S)} < 4.5~ {\rm GeV}$~\cite{belle1}
including the contributions of  $\sigma$ and $f_0(980)$. Removing the $\sigma$, the lower part of the spectrum gives a sensibly worse fit.}
\label{fig:dipion}
\end{figure}

Finally we considered the $\pi\pi$ invariant mass spectrum as measured by Belle in the mass region of the $\yb$~\cite{belle2} (see Fig.~\ref{fig:dipion}). In absence of background (see Fig.~\ref{fig:tune}), we fit for the sum of two relativistic Breit-Wigner functions describing the $\sigma$  and $f_0(980)$. The resulting fitted masses and widths are   
$m_{\sigma} = (714 \pm 77) ~{\rm MeV} $, $\Gamma_{\sigma} = 
(499 \pm 176)~ {\rm MeV} $ and 
$m_{f_{0}} = (955.3 \pm 4.5) ~{\rm MeV} $, $\Gamma_{f_{0}} = (14 \pm 15) ~{\rm MeV}$, in agreement with the standard knowledge about the light scalar mesons.
Since these mesons are the best currently known candidates for tetraquarks~\cite{hooft}, the four-quark nature of the \yb\ is naturally suggested. 
Furthermore we can extract  the ratio
\begin{equation}
{\cal B}(\yb \to  \psi(2S)\sigma(600))/{\cal B}(\yb\to \psi(2S)f_0(980))=2.0\pm 0.3.
\label{scalar}
\end{equation}

{\bf \emph{Interpretation of $Y_B$}}.
The dominance of decay modes with two baryons or with scalar mesons are a strong indication of a 
 $[cq][\bar c \bar q]$ diquark-antidiquark structure of the $Y_B$ meson.
\begin{figure}
\begin{center}
\SetScale{0.5}
\SetOffset(135,10)
\fcolorbox{white}{white}{
   \begin{picture}(295,202) (131,-70)
    \SetWidth{1.0}
    \SetColor{Black}
    \Line(136,203)(136,99)
    \Line(226,202)(226,98)
    \Line(137,150)(226,150)
    \SetWidth{0.5}
    \Vertex(226,204){4.24}
    \Vertex(226,99){4.24}
    \SetWidth{1.0}
    \Line(331,203)(331,99)
    \SetWidth{0.5}
    \Vertex(332,203){4.24}
    \Vertex(332,101){4.24}
    \SetWidth{1.0}
    \Line(331,151)(361,151)
    \SetWidth{0.5}
    \Vertex(363,151){4.24}
    \Vertex(391,150){4.24}
    \SetWidth{1.0}
    \Line(391,150)(421,150)
    \Line(422,203)(422,99)
    \SetWidth{0.5}
    \Vertex(422,99){4.24}
    \Vertex(422,203){4.24}
    \Vertex(136,99){4.24}
    \Vertex(136,204){4.24}
    \SetWidth{1.0}
    \Line(136,-66)(136,-12)
    \Line(136,-12)(226,-12)
    \Line(226,-12)(226,39)
    \Line(136,38)(175,-6)
    \Line(184,-17)(225,-65)
    \SetWidth{0.5}
    \Vertex(137,38){4.24}
    \Vertex(226,39){4.24}
    \Vertex(226,-66){4.24}
    \Vertex(137,-65){4.24}
    \SetWidth{1.0}
    \Line(332,39)(423,39)
    \Line(331,8)(331,-66)
    \Line(332,-28)(419,-28)
    \Line(421,9)(421,-65)
    \SetWidth{0.5}
    \Vertex(331,39){4.24}
    \Vertex(422,39){4.24}
    \Vertex(421,9){4.24}
    \Vertex(331,9){4.24}
    \Vertex(331,-65){4.24}
    \Vertex(421,-66){4.24}
    \Text(68,112)[]{$c$}
     \Text(113,112)[]{$\bar c$}
     \Text(167,112)[]{$c$}
     \Text(212,112)[]{$\bar c$}
     \Text(68,28)[]{$c$}
     \Text(113,28)[]{$\bar c$}
     \Text(167,28)[]{$c$}
     \Text(212,28)[]{$\bar c$}
     \Text(68,40)[]{$q$}
     \Text(113,40)[]{$\bar q$}
     \Text(167,40)[]{$q$}
     \Text(212,40)[]{$\bar q$}
     \Text(68,-43)[]{$q$}
     \Text(113,-43)[]{$\bar q$}
     \Text(167,-43)[]{$q$}
     \Text(212,-43)[]{$\bar q$}
      \Text(183,85)[]{$q^\prime$}
     \Text(195,85)[]{$\bar q^\prime$}
      \Text(176,5)[]{$s$}
     \Text(202,5)[]{$\bar s$}
      \Text(50,75)[]{${\bf (A)}$}
     \Text(230,75)[]{${\bf (B)}$}
      \Text(50,75)[]{${\bf (A)}$}
     \Text(230,75)[]{${\bf (B)}$}
      \Text(50,-7)[]{${\bf (C)}$}
     \Text(230,-7)[]{${\bf (D)}$}
  \end{picture}
}
\end{center}
\vspace{-1truecm}
\caption{Phase space allows ${\bf A}\to {\bf B} \;(Y\to \Lambda_c\bar \Lambda_c)$,  ${\bf A}\to {\bf C} \;(Y\to D\bar D^*) $,  ${\bf A}\to {\bf D} \;(Y\to 
\psi(2S) f_0(980)) $
}
\label{uno}
\end{figure}
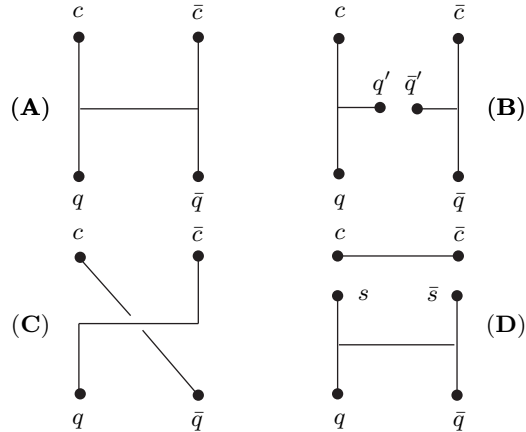
The diagram ${\bf A}$ in Fig.~\ref{uno} is a representation of a 
$[cq][\bar c \bar q]$ tetraquark where the two diquarks are 
connected by a string neutralizing their color.
Both diquarks have positive parity in the `single mode configuration', 
where the constituent quarks are unexcited with respect to one another. A $J^{PC}=1^{--}$ state must therefore have odd values of $\ell$.
The $\ell=1$ centrifugal potential favors configurations with the diquarks at an higher average relative distance with respect to the $\ell=0$ case, 
stretching  the color string which eventually breaks as in diagram ${\bf B}$ of Fig.~\ref{uno} (baryon-antibaryon decay). Breaking the color string in two points rather than in one, as shown in ${\bf C}$  or in ${\bf D}$  in Fig.~\ref{uno}, 
allows either the $D\bar D^{*}$ or the $\psi f_0$ decay.  
Actually also $\ell=3$ would be possible assuming the total spin 
of the diquarks to be $S=2$ with $S=S_\qq+S_{\bar \qq}$.  Although an analysis of the mass spectrum made along the lines 
of~\cite{dre2} would exclude this possibility, we will show that  a different implementation
 of the orbital angular momentum dependency of the Hamiltonian would still allow it. 
We aim  at estimating the mass of the $Y_B$ using a non-relativistic 
constituent quark model combined with a hadron string model. Using the fitted masses of constituent diquarks we will  find the spectrum as a function of the 
orbital angular momentum and radial excitation quantum numbers. 

As for the isospin assignment of the tetraquark wavefunction the observed decays
seem to favor $I=0$. 
On the other hand at the 
mass scale of the charm quark we could assume that light quark annihilation diagrams are suppressed so that the mass eigenvectors  align to the quark 
mass basis~\cite{rossivenez}. In the latter case one could have isospin 
breaking effects as is the case for the $X(3872)$ which turns out to decay 
in $J/\psi\, \rho$ and $J/\psi\, \omega$ at almost the same rate.
In a tetraquark interpretation one could explain such decay dynamics making Ê
the hypothesis that there are two states~\cite{prlmv} of the kind 
$X_u=[cu][\bar c\bar u]$ and $X_d=[cd][\bar c\bar d]$ with a difference in mass of 
the order of $7\times \cos(2\theta)$ MeV, $\theta$ being the mixing 
between $X_u$ and $X_d$ ($|\theta|\sim20^\circ$ in the case of $X(3872)$). 
From available data we still do not know if this 
is the case for the $Y_B$ particle although this is not affecting the 
following considerations.

{\bf \emph{Angular excitations}}.
The tetraquark picture with two diquarks connected by a color string allows to estimate the mass of the $Y_B$ and its orbital angular momentum excitation level. 
Considering the generalization of the Chew-Frautschi formula discussed in
Selem and Wilczek~\cite{wilc} we find in the mass limit of $M \omega$ ($M=m_1=m_2$) very large with respect to $\sigma/2\pi$~\footnote{We perform an infinite mass limit of the
the equations of a rotating relativistic string with tension $T=\sigma/2\pi$ and with two masses $m_1,m_2$ attached at its ends, namely~\cite{huang} 
\begin{eqnarray}
&&{\cal E}=\frac{m_1}{\sqrt{1-(\omega r_1)^2}}+\frac{m_2}{\sqrt{1-(\omega r_2)^2}}+\frac{\sigma}{2\pi \omega}\int_0^{\omega r_1} \frac{dv}{\sqrt{1-v^2}}+ \frac{\sigma}{2\pi \omega}\int_0^{\omega r_2} \frac{dv}{\sqrt{1-v^2}}\notag\\
&&\ell=\frac{\omega r_1^2 m_1}{\sqrt{1-(\omega r_1)^2}}+\frac{\omega r_2^2 m_2}{\sqrt{1-(\omega r_2)^2}} +\frac{\sigma}{2\pi \omega^2}\int_0^{\omega r_1} \frac{dv\; v^2}{\sqrt{1-v^2}}+ \frac{\sigma}{2\pi \omega^2}\int_0^{\omega r_2} \frac{dv\; v^2}{\sqrt{1-v^2}}\notag
\end{eqnarray}
}
\begin{equation} 
{\cal E}\approx 2 M+\frac{3}{(16\pi^2 M)^{1/3}} (\sigma \ell)^{2/3}
\label{eq:noi}
\end{equation}
where ${\cal E}$ is the energy of a system made up of two diquarks of mass $M=m_{[cq]}$ held together by a string characterized by a tension 
$ \sigma$ and spinning with some angular momentum $\ell$. The higher $\ell$ the better this model works as shown in Regge trajectories phenomenology. 
Taking $\sigma$, fitted from Regge slopes of non-charmed hadrons, to be $\sigma\sim 1$~GeV$^2$ and $M= 1933$~MeV~\cite{maiani},
we have ${\cal E}\approx 4311$~MeV for $\ell=1$ and ${\cal E}\approx 4793$~MeV for $\ell=3$. This means that if spin-spin interactions are neglected, which is a good approximation at $\ell\neq 0$, and if we consider no radial excitation, the $\ell=3$ assignation might be favored while the $\ell=1$ state would better match the $Y(4350)$.

To check the effectiveness of Eq.~(\ref{eq:noi}) in estimating the mass of known charmonium resonances having different $\ell$ it is worth noting that, for $M=1500$~MeV,  we
get $3500$~MeV for the average mass value of the $\chi_{cJ}$ system (experimentally located at $\sim 3493$~MeV), which has $\ell=1$, and $3770$~MeV for the $\ell=2$  case consistent with the $D$-wave nature of the $\psi(3770)$~\cite{faccini}. 
The fine structure of the $\chi_{cJ}$ system, with levels shifted from the average mass value, is due to spin-orbit and tensor interactions.

{\bf \emph{Radial excitations}}.
Let us assume that $Y_B$ is a diquark-antidiquark bound state 
in a Charmoinium-like potential. The difference in mass between the $n_r=0$ and the first radial excitation $n_r=1$ can be extracted from the available data on $\chi_{bJ}(2P),\chi_{bJ}(1P)$ states.
The mass difference $\Delta M=M_{\chi_b(2P)}-M_{\chi_b(1P)}\simeq 360$~MeV is expected to be roughly the same 
for charmed states, where instead no data are available for $\ell=1$ states. Therefore, neglecting spin-orbit terms and spin-spin interactions, 
the $(n_r=1,\ell=1)$ state would be located at $4671$~MeV, remarkably close to  $Y_B(4660)$, whereas  the  $(n_r=0,\ell=1)$ state 
would be located at $4311$~MeV, close to the $Y(4350)$.
We can make an estimate of the mass spectrum also using the constituent quark model discussed in~\cite{dre1}. If we neglect spin-spin and spin-orbit splittings and assume $\ell=1$ we 
find $M_{Y}=4361$~MeV if no radial excitation is taken into account, and $M_{Y_B}=4721$~MeV if $n_r=1$. Therefore we might conclude that $Y_B(4660)=Y(n_r=1,\ell=1)$  
and  $Y(4350)=Y(n_r=0,\ell=1)$.

{\bf \emph{Decays}}.
To refine the model we investigate whether the decay patterns of the $Y(4350)$ and $Y_B(4660)$ are consistent with the assigned quantum numbers:
from data we see that both $Y(4350)$ and $Y_B(4660)$ prefer to decay in $\psi (2S)\pi\pi$ rather than in  $J/\psi \pi\pi$ even if the latter mode is 
phase space enhanced.  The measured spectra show that the $\pi\pi$ pair comes from phase space (or, equivalently a $\sigma$) apart from the case of the 
$Y_B(4660)$ where there seems to be a 30\% component due to $f_0(980)$ (see Eq.~2).
We therefore attempt an explanation of the observed $Y$'s decay pattern describing the $S$-wave transition $\langle \psi (1S,2S) \mathbbmss{a} |Y\rangle$ as
\begin{equation}
\langle \psi(\eta,q) \mathbbmss{a}(k) |Y(\epsilon,p)\rangle =g \; \epsilon\cdot \eta
\label{matr0}
\end{equation}
where $ \mathbbmss{a}=\sigma,f_0$. 
We find  the eigenfunctions $\Psi$ of the linear part of the Cornell potential $V(r)=-k/r + r/a^2$ ($k=0.52$, $a=2.3$~GeV$^{-1}$) for the $\psi(1S,2S)$ and the charmonium-like $Y$, namely $\Psi=Y_{\ell,m}R_{n_r,\ell}$.
Interpreting the $Y$'s as  charmonium-like bound states made up by a diquark and an antidiquark we can write 
\begin{equation}
 g\propto \int d^3 r    \;R^{\dagger}_{{(\psi)}}(r)
R_{_{(Y)}}(r)
\label{matr}
\end{equation}
We are interested in estimating the following ratio of decay widths
\begin{equation}
\Gamma_\mathbbmss{a} (Y) \equiv \Gamma(Y\to  \psi(1S) (\pi\pi)_\mathbbmss{a} )/\Gamma(Y\to  \psi(2S) (\pi\pi)_\mathbbmss{a}) 
\end{equation} 
We find the values reported in the following Table (the square of the matrix element in Eq.~(\ref{matr}) weights the  three-body phase space where the two pions are the decay products of an intermediate~\footnote{The Breit-Wigner Ansatz we use is a rough approximation for the description of a very broad  $\sigma$ meson whereas is rather suitable for the $f_0(980)$.} scalar resonance; we include only those quantum number combinations which provide mass values close to the experimental ones)
\begin{center}
\begin{tabular}{|p{1.5cm}|p{1.5cm}|c|c|c||}
\hline
$n_{r}$ & $\mathbf{\ell}$ &   $\Gamma_{\sigma}(Y_B(4660))$  & $\Gamma_{f_0}(Y_B(4660))$\\
\hline 
$1$ & $1$ & $0.07$&$ 0.4$\\
\hline
$0$ & $3$ &$0.2$ &$1.2$\\
\hline
\hline
$n_{r}$ & $\mathbf{\ell}$ &  $\Gamma_{\sigma}(Y(4350))$& $\Gamma_{f_0}(Y(4350))$\\
\hline 
$0$ & $1$ &$3.5 $&-- \\
\hline
\end{tabular}
\end{center}
We report separately the $\sigma$ and $f_0$ contributions.
The results show that the radial excitation would explain
the preference of $Y_B$ tetraquark  to decay into $\psi(2S)\sigma$ whereas with this simple model we cannot explain why the  $Y(4350)$ seems to prefer the $\psi(2S)$ channel. This aspect remains puzzling.

{\bf \emph{Conclusions}}. 
We  reanalized the published data on the $Y(4630)$, $Y(4660)$ and $Y(4350)$ and obtained that the first two resonances correspond to a single state, 
the $Y_B(4660)$.
The large baryon decay rate and the decay into $\psi(2S) f_0(980)$ make the $Y_B$ an excellent candidate for a $[cd][\bar c \bar d]$ diquark-antidiquark bound state suggesting that $\yb$ represents  the first observation of a what we might call a {\it charmed baryonium}. The fact that the mass of the $Y_B$ is sensibly  higher 
than the baryon-antibaryon threshold excludes any baryon molecule picture.
Assuming that the interaction between the two diquarks has the same potential as that between quarks in the charmonium we estimate the mass spectra for the ground state and the first radial excitation for two values of $\ell$ and find that $Y_B$ best fits the first radial excitation of  $\ell=1$. With the same assumptions we show that the $Y(4350)$ fits the corresponding radial ground state and we show that  the absence of the decay into $J/\psi \pi\pi$ of the $Y_B(4660)$ seems compatible with its tetraquark interpretation.

\begin{acknowledgments}
\end{acknowledgments}
We wish to thank Gino Isidori, Antonio Vairo, and Christoph Hanhart  for comments and suggestions on the manuscript. 

\bigskip 

\end{document}